# How to Compute Halting

Eric C.R. Hehner

Department of Computer Science, University of Toronto
hehner@cs.utoronto.ca

**Abstract:** A consistently specified halting function may be computed.

## Halting Problem

Here is the function header and specification of a Pascal function *halts* to compute the termination status of Pascal procedures; the function body is absent. Following that, we have a Pascal procedure *diag* in its entirety.

**function** *halts* (*p*, *i*: **string**): **string**;
{ return 'yes' if *p* represents a Pascal procedure with one string input parameter }
{         whose execution terminates when given input *i* ; }
{ return 'no' if *p* represents a Pascal procedure with one string input parameter }
{         whose execution does not terminate when given input *i* ; }
{ return 'not applicable' if *p* does not represent a Pascal procedure }
{         with one string input parameter }

**procedure** *diag* (*s*: **string**);
**begin**
        **if** *halts* (*s*, *s*) = 'yes' **then** *diag* (*s*)
**end**

We assume there is a dictionary of function and procedure definitions that is accessible to *halts* , so that the call *halts* ('*diag*', '*diag*') allows *halts* to look up '*diag*' , and subsequently '*halts*' , in the dictionary, and retrieve their texts for analysis. What should the result of *halts* ('*diag*', '*diag*') be? This is a question about the specification of *halts* . Let's look at each possibility in turn.

Should the result of *halts* ('*diag*', '*diag*') be 'not applicable' ? Syntactically, *diag* is a procedure; to determine that *halts* is being used correctly within *diag* , we need only the header for *halts* , not the body, and we have the header. Semantically, it is a procedure; to determine the meaning of the call to *halts* within *diag* , we need only the specification of *halts* , not its implementation, and we have the specification. (That important programming principle enables a programmer to call procedures written by other people, knowing only the specification, not the implementation. It also enables a programmer to change the implementation of a procedure, but still satisfying the specification, without knowing where and why the procedure is being called.) So there is nothing wrong with the definition of *diag* , and the result should not be 'not applicable' .

Should the result of *halts* ('*diag*', '*diag*') be 'yes' ? If so, the semantics of *diag* ('*diag*') is nontermination, so it should be 'no' .

Should the result of *halts* ('*diag*', '*diag*') be 'no' ? If so, the semantics of *diag* ('*diag*') is termination, so it should be 'yes' .

We have ruled out all possibilities. Therefore the *halts* specification is inconsistent, and *halts* cannot be programmed according to its specification.

## How to Compute Limited Halting

It is inconsistent to ask for a Pascal function to compute the halting status of all Pascal procedures. But we can ask for a Pascal function to compute the halting status of some Pascal procedures. For example, a function to compute the halting status of just the two procedures

**procedure** *loop* (*s*: **string**); **begin** *loop* (*s*) **end**
**procedure** *stop* (*s*: **string**); **begin end**

is easy. Perhaps we can ask for a Pascal function to compute the halting status of all Pascal procedures that do not refer to this halting function, neither directly nor indirectly. Here is its header, specification, and a start on its implementation.

**function** *halts1* (*p*, *i*: **string**): **string**;
{ return 'yes' if *p* represents a Pascal procedure with one string input parameter }
{       that does not refer to *halts1* (neither directly nor indirectly) }
{       and whose execution terminates when given input *i* ; }
{ return 'no' if *p* represents a Pascal procedure with one string input parameter }
{       that does not refer to *halts1* (neither directly nor indirectly) }
{       and whose execution does not terminate when given input *i* ; }
{ return 'maybe' if *p* represents a Pascal procedure with one string input parameter }
{       that refers to *halts1* (either directly or indirectly); }
{ return 'not applicable' if *p* does not represent a Pascal procedure }
{       with one string input parameter }
**begin**
        **if** ( *p* does not represent a Pascal procedure with one string input parameter)
        **then** *halts1*:= 'not applicable'
        **else if** ( *p* refers to *halts1* directly or indirectly)
            **then** *halts1*:= 'maybe'
            **else** (return halting status of *p* , either 'yes' or 'no' )
**end**

The first case checks whether *p* represents a (valid) procedure exactly as a Pascal compiler does. The middle case looks like a transitive closure algorithm, but it is problematic because, theoretically, there can be an infinite chain of calls. Thus we may be able to compute halting for this limited set of procedures, but not determine whether a procedure is in this limited set. The last case may not be easy, but at least it is free of the reason it has been called incomputable: that it cannot cope with

**procedure** *diag1* (*s*: **string**);
**begin**
        **if** *halts1* (*s*, *s*) = 'yes' **then** *diag1* (*s*)
**end**

Procedure *diag1* refers to *halts1* by calling it, so *halts1* is not required to determine the halting status of *diag1* . Therefore *halts1* ('*diag1*', '*diag1*') = 'maybe' , and execution of *diag1* ('*diag1*') is terminating.

Calling is one kind of referring, but not the only kind. In the specification of *halts1* , the name *halts1* appears, and also in the body. These are self-references, whether or not *halts1* calls itself. We exempt *halts1* from having to determine the halting status of procedures containing any form of reference to *halts1* ; the result is 'maybe' . We might try to circumvent the limitation by writing another function *halts2* that is identical to *halts1* but renamed (including in the specification, the return statements, and any recursive calls).

```
function halts2 (p, i: string): string;
{ return 'yes' if p represents a Pascal procedure with one string input parameter }
{      that does not refer to halts2 (neither directly nor indirectly) }
{      and whose execution terminates when given input i ; }
{ return 'no' if p represents a Pascal procedure with one string input parameter }
{      that does not refer to halts2 (neither directly nor indirectly) }
{      and whose execution does not terminate when given input i ; }
{ return 'maybe' if p represents a Pascal procedure with one string input parameter }
{      that refers to halts2 (either directly or indirectly); }
{ return 'not applicable' if p does not represent a Pascal procedure }
{      with one string input parameter }
begin
       if ( p does not represent a Pascal procedure with one string input parameter)
       then halts2:= 'not applicable'
       else if ( p refers to halts2 directly or indirectly)
            then halts2:= 'maybe'
            else (return halting status of p , either 'yes' or 'no' )
end;
```

Of course, *halts2* has its own nemesis:

```
procedure diag2 (s: string);
begin
       if halts2 (s, s) = 'yes' then diag2 (s)
end
```

The point is that *halts2* can determine halting for procedures that *halts1* cannot, and *halts1* can determine halting for procedures that *halts2* cannot. For example,

*halts1* ('*diag1*', '*diag1*') = 'maybe'   because *diag1* calls *halts1*
*halts2* ('*diag1*', '*diag1*') = 'yes'     because execution of *diag1* ('*diag1*') terminates
*halts2* ('*diag2*', '*diag2*') = 'maybe'   because *diag2* calls *halts2*
*halts1* ('*diag2*', '*diag2*') = 'yes'     because execution of *diag2* ('*diag2*') terminates

But there are procedures that refer to both *halts1* and *halts2* , for which both *halts1* and *halts2* say 'maybe' . The most interesting point is this: even though *halts1* and *halts2* are identical except for renaming, they produce different results when given the same input, according to their specifications.

## How to Compute Unlimited Halting

In Pascal, as originally defined, identifiers cannot contain underscores. I now define a new programming language, Pascal_, which is identical to Pascal except that identifiers can contain underscores. Pascal_ is a larger language than Pascal, but no more powerful: they are both Turing-Machine-equivalent. In this new language, perhaps we can write a function named *halts_* that determines the halting status of all Pascal procedures. Pascal procedures are syntactically prevented from referring to *halts_* , so the problem of determining whether a Pascal procedure refers to *halts_* disappears, along with the 'maybe' option.

```
function halts_ (p, i: string): string;
{ return 'yes' if p represents a Pascal procedure with one string input parameter }
{      whose execution terminates when given input i ; }
{ return 'no' if p represents a Pascal procedure with one string input parameter }
{      whose execution does not terminate when given input i ; }
{ return 'not applicable' if p does not represent a Pascal procedure }
{      with one string input parameter }
```

**begin**
    **if** ( $p$  does not represent a Pascal procedure with one string input parameter)
    **then** *halts_* := 'not applicable'
    **else** (return halting status of  $p$ , either  'yes'  or  'no' )
**end**

If it is possible to write a Pascal function to compute the halting status of all Pascal procedures that do not refer to this halting function, then by writing in another language, we can compute the halting status of all Pascal procedures.

There is an argument that, at first sight, seems to refute the possibility of computing the halting status of all Pascal procedures just by programming in another language.  Suppose that in writing  *halts_*  we do not use any underscores in any other identifiers, and we do not use the identifier  *halts* .  Then we can easily obtain a Pascal function  *halts*  just by deleting the underscore from the  *halts_*  identifier.  We thus obtain a Pascal function with the same functionality: *halts* ($p$, $i$) = *halts_* ($p$, $i$)  for all  $p$  and  $i$ .  But there cannot be a Pascal function that computes the halting status of all Pascal procedures.  Therefore, the argument concludes, there cannot be a Pascal_ function to do so either.

As compelling as the previous paragraph may seem, it is wrong.  We have already seen that renaming *halts1*  to  *halts2*  produces a function with different results.  The phenomenon can be understood in everyday experience.  If I say "My name is Eric Hehner." I am telling the truth. If Margaret Jackson says exactly the same words, she is lying.  When I say it, there is a self-reference;  when Margaret Jackson says it, there is no self-reference.  The truth of that sentence depends on who says it.

Here is a simple example of the failure of program translation:  a Pascal_ procedure that prints its own name.

**procedure** *A_*;  { this procedure prints its own name }
**begin** *print* ('A_') **end**

Translating this procedure to Pascal, we face a dilemma.  We could translate it as

**procedure** *A*;  { this procedure prints its own name }
**begin** *print* ('A_') **end**

arguing that the two procedures have the same output, but clearly this translation does not preserve the intention.  The Pascal_ procedure  *A_*  meets its specification;  the Pascal translation  *A*  does not.  Or we could translate it as

**procedure** *A*;  { this procedure prints its own name }
**begin** *print* ('A') **end**

arguing that we have preserved the intention, but clearly the two procedures do not have the same output. Translating from  *halts_*  to  *halts*  has the same problem.  We cannot preserve the intention because the specification at the head of  *halts_* , which is perfectly reasonable for a Pascal_ function, becomes inconsistent when placed at the head of a Pascal function.  If we just use the same Pascal_ procedure but delete the underscores, we obtain a Pascal procedure that no longer satisfies the specification.

There is another argument that, at first sight, also seems to refute the possibility of computing the halting status of all Pascal procedures just by programming in another language.  In Pascal, we can write an interpreter for Pascal_ programs.  So if we could write a halting function *halts_* in Pascal_ for all of Pascal, we could feed the text of *halts_* to this interpreter, and thus obtain a Pascal function to compute halting for all Pascal procedures. But there cannot be a Pascal function that computes the halting status of all Pascal procedures. Therefore, the argument concludes, there cannot be a Pascal_ function to do so either.

The reason this argument fails is the same as the reason the previous argument fails. The interpreter interpreting *halts_* is just like the translation of *halts_* into Pascal by deleting underscores. The interpreter interpreting *halts_* can be called by another Pascal program; *halts_* cannot be called by a Pascal program. That fact materially affects their behavior. Pascal_ program *halts_* can be applied to a Pascal procedure *d* that calls the interpreter interpreting *halts_* applied to *d*, and it will produce the right answer. But the interpreter interpreting *halts_* applied to *d* calls the interpreter interpreting *halts_* applied to *d*, and execution will not terminate.

### the Barber

A town named Russellville consists of some men (only men). Some of the men shave themselves; the others do not shave themselves. A barber is a person who shaves all and only those men in Russellville who do not shave themselves. There is a barber: a man named Bertrand_ who lives just outside the town, in the Greater Russellville Area called Russellville_. Without any difficulty, he satisfies the specification of barber.

One of the men in Russellville, whose name is Bertrand, decided that there is no need to bring in a barber from outside town. Bertrand decided that he could do the job. He would shave those men whom Bertrand_ shaves, and not shave those men whom Bertrand_ does not shave. If Bertrand_ is fulfilling the role of barber, then by doing exactly the same actions as Bertrand_, Bertrand reasoned that he would fulfill the role of barber. But Bertrand is wrong; those same actions will not fulfill the role of barber when Bertrand performs them. To be a barber, Bertrand has to shave himself if and only if he does not shave himself. A specification that is perfectly consistent and possible for someone outside town becomes inconsistent and impossible when it has to be performed by someone in town.

And so it is with the halting specification, and for the same reason. For Bertrand_, the barber specification has no self-reference; for Bertrand, the barber specification has a self-reference. For *halts_*, the halting specification has no self-reference; for *halts*, the halting specification has a self-reference (indirectly through *diag* and other procedures that call *halts*).

### Conclusion

By weakening the specification a little, reducing the domain from "all procedures" to "all procedures that do not refer to the halting function", we obtain a specification that may be both consistent and computable. Equivalently, we may be able to compute the halting status of all procedures in a Turing-Machine-equivalent language by writing a halting function in another Turing-Machine-equivalent language, assuming that the procedures of the first language cannot refer to the halting function written in the second language. In any case, we do not yet have a proof that it is impossible.

### Reference

A.M.Turing: on Computable Numbers with an Application to the Entscheidungsproblem, *Proceedings of the London Mathematical Society* s.2 v.42 p.230-265, 1936; correction s.2 v.43 p.544-546, 1937

[other papers on halting](#)